\documentclass[a4paper,11pt]{article}
\pdfoutput=1 % if your are submitting a pdflatex (i.e. if you have
\usepackage{jheppub} % for details on the use of the package, please
\usepackage[T1]{fontenc} % if needed

\usepackage{stackrel}
\usepackage{tikz-cd} 
\usepackage{multicol}
\usepackage{multirow}

\newcommand{\eq}{\begin{equation}}
\newcommand{\eqx}{\end{equation}}
\newcommand{\eqn}{\begin{eqnarray}}
\newcommand{\eqnx}{\end{eqnarray}}
\newcommand{\nn}{\nonumber}
\newcommand{\vn}{\vec{n}}
\newcommand{\ve}{\vec{e}}
\newcommand{\cN}{{\cal N}}

\usepackage{currvita}

\title{\boldmath Constraints of kinematic bosonization in two and higher dimensions}

\author[a]{A. Bochniak,}
\author[a]{B. Ruba,}
\author[a,1]{J. Wosiek\note{Corresponding author.}}
\author[a]{and A. Wyrzykowski}

\affiliation[a]{Institute of Theoretical Physics, Jagiellonian University, Poland}

\emailAdd{arkadiusz.bochniak@doctoral.uj.edu.pl}
\emailAdd{blazej.ruba@doctoral.uj.edu.pl}
\emailAdd{jacek.wosiek@uj.edu.pl}
\emailAdd{adwyrzykowski@gmail.com}

\abstract{Despite being less known, local bosonizations of fermionic systems exist in spatial dimensions higher than one. Interestingly, the dual bosonic systems are subject to local constraints, as in theories with gauge freedom. These constraints effectively implement long distance exchange interactions. In this work we study in detail one such system, proposed a long time ago. Properties of the constraints are elaborated for two-dimensional, rectangular lattices of arbitrary sizes. For several small systems the constraints are solved analytically. It is checked that spectra of reduced spin hamiltonians agree with the original fermionic ones. The equivalence is extended to fermions in presence of background Wegner $\mathbb{Z}_2$ fields coupling to fermionic parity. This is illustrated by an explicit calculation for a particular configuration of Wegner's variables. Finally, a possible connection with the recently proposed web of dualities is discussed.
}

\begin{document} 
\maketitle
\flushbottom

\section{Introduction}
\label{sec:intro}

Relation between fermionic and spin degrees of freedom is an old subject \cite{JW,YN}, but it still attracts a fair amount of interest. There is a variety of motivations for such studies. Presence of Grassmann variables in fermionic field theories leads to practical difficulties in their study, hence the desire to eliminate them \cite{JBK,JAW}. Secondly, equivalences between apparently different physical systems often offer new insights into their dynamics. There has been a lot of progress in these directions recently. For instance, it has been shown \cite{CKR,YAC} that fermions in space dimension $d$ can be exactly mapped to a local generalized gauge theory on the dual lattice, with $\mathbb Z_2$ gauge variables associated to $(d-1)$-dimensional objects (hence an Ising model for $d=1$, standard gauge theory with modified Gauss' law for $d=2$ and so-called higher gauge theories for $d \geq 3$). This idea has been motivated by studies of fermions in topological quantum field theories \cite{GK}. There exists also a variety of known dualities in the continuum, especially in low dimensions \cite{KT, SSWX,SSWW}. Many of them have been discovered in string theoretic considerations. Some of them connect bosons to fermions, which provides another point of view on bosonizaton. Finally, intensive studies of quantum computers and "quantum algorithms" stimulate some progress in the hamiltonian formulation, see in particular \cite{K,BK,ZC, K1}.

Spin-fermion maps are particularly well understood and exploited in systems of spatial dimension one. Their extensions to higher dimensions typically lead to complicated non-local interactions or constraints and seems to be not practical.

In this paper we revisit an old proposal \cite{JAW,SZ} in which spins interact locally and satisfy local constraints. These constraints effectively take care of the non-locality of fermions in arbitrary space dimensions. 

Let us begin with a simple fermionic hamiltonian on a one dimensional lattice
\eqn
H_f= i \sum_n \left(\phi(n)^{\dagger}\phi(n+1) - \phi(n+1)^{\dagger}\phi(n)\right) ,\;\;\;\;\{\phi(m)^{\dagger},\phi(n)\}=\delta_{m n}. \label{hf}
\eqnx
Its equivalent in terms of spin variables reads
\eqn
H_s= \frac{1}{2} \sum_n \left(\sigma^1(n)\sigma^2(n+1) - \sigma^2(n)\sigma^1(n+1)\right),   \label{hspin1}
\eqnx
where Pauli matrices $\sigma^k(n)$ commute between different sites labelled by $n$. Boundary conditions for $\sigma^1$ and $\sigma^2$ are taken to be opposite to (resp. the same as) boundary conditions for fermions if the number of fermions $\sum\limits_{n} \phi(n)^{\dagger} \phi(n)$ is even (resp. odd). The standard way to derive this equivalence is via the Jordan-Wigner transformation \cite{JW}. Direct generalization of this method to higher dimensions leads to non-local spin-spin interactions. Therefore we adopt another route, which applies also to multidimensional systems.

To this end we introduce the following Clifford variables (also called Majorana fermions)
\eqn
X(n)=\phi(n)^{\dagger}+\phi(n),&&  \;\;\;\; Y(n)=i(\phi(n)^{\dagger}-\phi(n)),\label{cf}
\eqnx
and rewrite the fermionic hamiltonian \eqref{hf} in terms of link (or hopping) operators
\eqn
H_s=\frac{1}{2} \sum_n  \left(S(n) + \widetilde{S}(n)\right), \\
S(n)=iX(n)X(n+1),&&\;\;\;\;\widetilde{S}(n)=iY(n)Y(n+1). \nn
\eqnx
Link operators obey the following relations
\begin{equation}
    \begin{split}
    S(n)^2&=1,\\
    [S(m), S(n)] = 0, &\qquad m\neq n-1,n+1, \\
     \{S(m),S(n)\} =0, &\qquad m=n-1,n+1.
    \end{split}
    \label{alg}
\end{equation}

In words, they square to one, anticommute if they share one common vertex and commute otherwise. Analogous relations hold also with $S$ replaced by $\widetilde S$ in the above. Furthermore $S$ and $\widetilde S$ always commute with each other:
\eqn
[ S(m), \widetilde{S}(n) ] =0. 
\label{com}
\eqnx

It can be shown that all relations in the algebra generated by $S$ and $\widetilde S$ operators follow from these already listed. Furthermore this algebra has only two irreducible representations, corresponding to two possible values of fermionic parity. Therefore in order to perform bosonization it is sufficient to construct operators obeying relations \label{alg,com} in terms of spin operators. One such representation reads
\eq
S(n)=\sigma^1(n)\sigma^2(n+1),\qquad\widetilde{S}(n)=-\sigma^2(n)\sigma^1(n+1).
\eqx
Replacing operators $S(n)$ in the spin hamiltonian by their spin representatives gives \eqref{hspin1}. 

In this way we have changed fermionic and spin variables without invoking the Jordan-Wigner transformation. This lends itself an interesting possibility that similar construction exists in higher dimensions.

Before concluding this Section we note that at the heart of the equivalence claim is the meta-principle that systems described by the same algebras of operators are equivalent. One concrete substantiation of this, relevant for representations of Heisenberg groups, is~given by the celebrated Stone-von Neumann theorem \cite{SvN}. See \cite{SZ,CKR,BBR} for discussion of this for algebras of fermionic bilinears, which are directly relevant for the present work.

All systems discussed in this work are defined on finite lattices. This leads to an interesting interplay between boundary conditions, conserved charges and constraints. Explanation of these issues is one of the goals of the present paper.

In the next Section we review the spin-fermion correspondence in spatial dimension two, including the definition of constraints present in this model. In Section \ref{sec:constraints} we explain the interplay between boundary conditions and fermionic parity. Furthermore we solve the constraints for few small systems and check explicitly that the spectra of fermionic and spin hamiltonians do coincide. In Section \ref{sec:generalization} we show that constraints can be interpreted as the condition that certain $\mathbb Z_2$ gauge field hidden in the bosonic theory is trivial. Modifying the form of constraints is equivalent to coupling fermions to an external gauge field. This is illustrated by a concrete calculation, in which fermions in a constant magnetic field are considered. We conclude in Section \ref{sec:summary} and discuss a very attractive potential relation with the rapidly developing family of dualities in ($2+1$) dimensions. 

\section{The equivalent spin model in two dimensions}
\label{sec:equivalent_spin_model}

Generalization of the above idea to two and higher space dimensions is known for a long time \cite{JAW}. In two dimensions the fermionic hamiltonian
\eqn
H_f= i \sum_{\vn,\ve} \left(\phi(\vn)^{\dagger}\phi(\vn+\ve) - \phi(\vn+\ve)^{\dagger}\phi(\vn)\right) =\frac{1}{2} \sum_l  \left(S(l) + \widetilde{S}(l)\right),\;\;\;l=(\vn,\ve) \label{hf2}
\eqnx
can be again rewritten in terms of two types of hopping operators labelled by links of a two dimensional lattice. They obey relations which are a straightforward generalization of these from the one dimensional case. In short: the hopping operators of the same type commute unless corresponding links have one common site. The difference is that now four, instead of two anticommuting link operators, are attached to each lattice site. Consequently, one needs bigger matrices to satisfy the corresponding algebra in higher dimensions.

\begin{figure}[h]
\begin{center}
    \begin{tikzcd}
    {}&{}\arrow[d,-,ultra thick] &{}\arrow[d,-,ultra thick] &{}\arrow[d,-,ultra thick] & {}\\
  {}\arrow[r,-,ultra thick] &{}{\scriptstyle \bold 3}\stackbin[\bold 4]{\bold 2}{\bullet}{\scriptstyle \bold 1} \arrow[d,-,ultra thick]\arrow[r,-,ultra thick] & {}{\scriptstyle \bold 3}\stackbin[\bold 4]{\bold 2}{\bullet}{\scriptstyle \bold 1} \arrow[r,-,ultra thick] \arrow[d,-,ultra thick]& {}{\scriptstyle \bold 3}\stackbin[\bold 4]{\bold 2}{\bullet}{\scriptstyle \bold 1} \arrow[r,-,ultra thick]\arrow[d,-,ultra thick]&{}  \\
 {}  \arrow[r,-,ultra thick]& {}{\scriptstyle \bold 3}\stackbin[\bold 4]{\bold 2}{\bullet}{\scriptstyle \bold 1}\arrow[d,-,ultra thick] &{}{}{\scriptstyle \bold 3}\stackbin[\bold 4]{\bold 2}{\bullet}{\scriptstyle \bold 1} \arrow[l,-,ultra thick]\arrow[r,-,ultra thick]\arrow[d,-,ultra thick] & {}{\scriptstyle \bold 3}\stackbin[\bold 4]{\bold 2}{\bullet}{\scriptstyle \bold 1} \arrow[r,-,ultra thick]\arrow[d,-,ultra thick]& {}\\
 {} \arrow[r,-,ultra thick] &{}{\scriptstyle \bold 3}\stackbin[\bold 4]{\bold 2}{\bullet}{\scriptstyle \bold 1}\arrow[d,-,ultra thick] &{}{\scriptstyle \bold 3}\stackbin[\bold 4]{\bold 2}{\bullet}{\scriptstyle \bold 1}\arrow[l,-,ultra thick]\arrow[r,-,ultra thick]\arrow[d,-,ultra thick] & {}{\scriptstyle \bold 3}\stackbin[\bold 4]{\bold 2}{\bullet}{\scriptstyle \bold 1} \arrow[r,-,ultra thick]\arrow[d,-,ultra thick]& {}\\
 {}&{}&{}&{}&{}
\end{tikzcd}
\end{center}
\caption{Assignment of the Dirac matrices to lattice vertices \eqref{lgam}.}
\label{fig:lattice_gamma}
\end{figure}
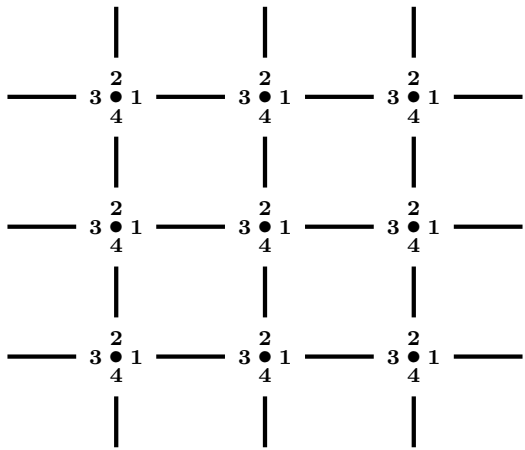

In two dimensions we choose Euclidean Dirac matrices and set (cf. Figure \ref{fig:lattice_gamma})
\eqn
S(\vec{n},\hat{x})=\Gamma^1(\vec{n})\Gamma^3(\vec{n}+\hat{x}),&\;\;\;\;\;&S(\vec{n},\hat{y})=\Gamma^2(\vec{n})\Gamma^4(\vec{n}+\hat{y}), \nn \\
\widetilde{S}(\vec{n},\hat{x})=\widetilde{\Gamma}^1(\vec{n})\widetilde{\Gamma}^3(\vec{n}+\hat{x}),&\;\;\;\;\;&\widetilde{S}(\vec{n},\hat{y})=\widetilde{\Gamma}^2(\vec{n})\widetilde{\Gamma}^4(\vec{n}+\hat{y}),  \label{lgam}
\eqnx
\eq
\widetilde{\Gamma}^k=i \prod\limits_{j\neq k} \Gamma^j.  \nn
\eqx 

It is a straightforward exercise to show that the two dimensional analogue of relations \eqref{alg} remain satisfied. Hence our hamiltonian in the spin representation reads
\eq
H_s=\frac{1}{2} \sum_l  \left(S(l) + \widetilde{S}(l)\right).   \label{hs}
\eqx

Generalization to higher dimensions is simple. One needs representations of higher Clifford algebra, i.e. by larger Dirac matrices. In $d$ dimensions we use $2d$ anticommuting ones which corresponds to the $2d$ links meeting at one lattice site.
Consequently, we~have a viable candidate for a local bosonic system equivalent to free fermions in arbitrary dimensions.

The story is not over, however, since representation \eqref{lgam} is redundant with respect to the fermionic one. In fact, in two space dimensions, it doubles the number of degrees of freedom per lattice site compared to the original fermionic system. Evidently one needs additional constraints for above spins to render the exact correspondence. This can be traced to the fact that original fermionic operators $S$ and $\widetilde S$ obey additional relations, not present in spatial dimension one. These will have to be imposed as constraints on physical states in the spin system.
 
Necessary constraints are provided by the plaquette operators $P_n$ (from now on $n$ is a~two dimensional index $n=(n_x, n_y)$). If we denote by $C_n$ an elementary plaquette labelled by its lower-left corner, say, then
 \eq
 P_n=\prod_{l  \in C_n }S(l). \label{cons}
 \eqx
These operators are identically $1$ in the fermionic representations, while in the spin representation they merely satisfy $P_{n}^2=1$. Hence imposing constraints
  \eq
  P_{n}=1 \label{con}
 \eqx
is necessary for the validity of the fermion-spin equivalence. It was shown already in \cite{JAW} that \eqref{con}  indeed correctly reduces the number of degrees of freedom per lattice site.  

Details of how the claimed reduction works depend on the lattice size, boundary conditions and other specifications. Detailed answer to this and related questions is the aim of the present work, as continued in the next sections. General explanations are given, with checks on small lattices performed analytically using symbolic algebra software \cite{MATH}.
 
\section{The constraints}
\label{sec:constraints}
The precise form of constraints that have to be imposed in order to make the above fermion-spin equivalence valid depends on the geometry of the lattice. To illustrate this feature we consider two-dimensional $L_x \times L_y$ rectangular lattices.\footnote{We assume $L_x,L_y\geqslant 3$ to avoid certain pathologies.} Periodic or antiperiodic boundary conditions are used. Different periodicity conditions for fermions and spins are allowed:
\eq
\phi(n+L_x \hat{x})=\epsilon_x \phi(n),\;\;\;\Gamma^k(n+L_x \hat{x})=\epsilon'_x\Gamma^k(n),\;\;\;\;\;\epsilon_x,\epsilon'_x=\pm 1,
\eqx
and similarly for the other direction.

We seek to impose ${\cal N}=L_x L_y$ constraints \eqref{con} to eliminate abundant degrees of freedom. However, not all of them are independent. For example, in the spin representation plaquette operators satisfy the identity
\eq
\prod_n P_n =1,
\eqx
which leaves at most ${\cal N}-1$ independent constraints.

In addition, on finite periodic lattices one can also construct "Polyakov line" operators 
\eq
{\cal L}_x(n_y)=\prod_{n_x=1}^{L_x} S(n_x,n_y,\hat{x}),\qquad{\cal L}_y(n_x)=\prod_{n_y=1}^{L_y} S(n_x,n_y,\hat{y}).
\eqx
In fermionic representation they are just pure numbers sensitive to the boundary conditions, while in spin representation their squares are unity, similarly to the plaquette operators. Hence again they provide additional projectors.
In principle there are $L_x+L_y$ line operators, but in fact they can be shifted perpendicularly by multiplying them with appropriate rows or columns of plaquette operators.\footnote{In fact one can even deform them to products of hopping operators along not necessarily straight lines. What really matters here is their winding number.}
Therefore, altogether there are only two more candidates for independent projectors.

It has been shown in \cite{SZ} that there are no further constraints that have to be imposed besides those defined by plaquette and line operators. This has also been revisited and generalized in the work \cite{BBR}, which was done in parallel to this paper.

It turns out that even this set of $\cN-1$ plaquettes and two line projectors is overcomplete. The additional  structure is revealed once we consider the operator of fermion number at each site (i.e. the fermion density) 
\eq
N(n)=\phi^\dagger(n)\phi(n).
\eqx
Since hamiltonian \eqref{hf2} is moving fermions between neighbouring sites only, the total number of fermions, $N=\sum\limits_n N(n)$, is conserved, but obviously their density $N(n)$ is not.

In the spin representation the number operator is related to the $\Gamma^5$ matrix 
\eq
\Gamma^5(n)=\eta (-1)^{N(n)}=\eta \left(1-2N(n)\right),
\eqx
where $\eta=\pm 1$ represents the freedom of defining a fermion-empty and a fermion-occupied state in the spin representation. In particular the total fermionic parity $(-1)^N$ is given by the product of $\Gamma^5(n)$ over all lattice sites. As in the fermionic representation, $N$ is conserved, while the number densities $N(n)$ are not.
On the other hand, the plaquette and line operators do commute with the local densities. This will be exploited below when we diagonalize constraints.

Calculating directly from the definition of $\mathcal L$ and $(-1)^N$ operators in the spin representation one obtains the following identity
\eq
\Pi \equiv \prod_{n_y=1}^{L_y} {\cal L}_x(n_y) \prod_{n_x=1}^{L_x} {\cal L}_y(n_x) =(-\epsilon'_x)^{L_y}(-\epsilon'_y)^{L_x} (-\eta)^{L_x L_y} (-1)^N, \label{prod}
\eqx
which (for fixed $(-1)^N$) implies a relation between two Polyakov line projectors if at least one of $L_x$, $L_y$ is odd. On the other hand, if both $L_x$ and $L_y$ are even, the left hand side is insensitive to the choice of one of two values of Polyakov lines. Indeed, on the subspace defined by plaquette constraints one has $\Pi = {\cal L}_x(n_y)^{L_y} {\cal L}_y(n_x)^{L_x}$, which is a c-number if $L_x$ and $L_y$ are even.  

Another crucial ingredient in understanding the structure of constraints is derived by evaluating the value of $\Pi$ in fermionic representation. Comparing with the result \eqref{prod} one obtains the identity
\eq
(-1)^N=\eta^{L_x L_y} \left(-\frac{\epsilon'_x}{\epsilon_x}\right)^{L_x} \left(-\frac{\epsilon'_y}{\epsilon_y}\right)^{L_y}.   \label{cond}
\eqx
This means that for given value of $\eta$, and boundary conditions for fermions and spins, only one of the two possible values of $(-1)^N$ is realized. This means that for the other there do not exist any solutions of constraints. We remark that a formula analogous to \eqref{cond} (though in general not as transparent) exists also for more general lattice geometries.

Recall that in the ordinary fermionic Fock space the dimension of the space of states for a given value of $(-1)^N$ is $2^{{\cal N} -1}$. On the other hand in our generalized system without constraints imposed this dimension is equal to $4^{{\cal N}-1}$, so it is too large by a factor $2^{{\cal N}}$. Thus it is natural to anticipate that there should be $\mathcal N$ independent constraints, each of which reduces the Hilbert space dimension by a factor of two. 

We have already shown that at most $\mathcal N -1$ plaquettes are independent and that if at least one of $L_x$ and $L_y$ is odd, then one Polyakov line can be eliminated in favor of the other constraints. Thus in this case one has exactly $\mathcal N-1$ independent plaquettes and one independent Polyakov loop. On the other hand if $L_x$ and $L_y$ are even, it is not possible to eliminate one of line operators. Hence it must be that only $\mathcal N-2$ plaquettes are independent. This is indeed the case, as will be explained in the Subsection \ref{sec:explicit_examples}.

It is now known \cite{SZ, BBR} that it is always possible in principle to find $2^{\mathcal N -1}$ linearly independent solutions of constraints, corresponding to $2^{\mathcal N-1}$ basis vectors in one half of the Fock space. Besides the restriction to a fixed value $(-1)^N$, the two systems are indeed equivalent: there exists a unitary operator between their Hilbert spaces which carries even (i.e.~ commuting with $(-1)^N$) operators to spin operators according to the presented prescription. In~particular any fermionic hamiltonian, which is always even, has the same spectrum in the fermionic representation and in the spin representation. 

On the other hand explicit solutions of constraints are known only in certain special cases. In the forthcoming discussion we will discuss how constraints can be solved, at least for small lattice sizes. Results of all these calculations, carried out using symbolic algebra software, are in accord with theoretical predictions outlined above, providing a solid check of correctness. Needless to say, development of practical ways to deal with constraints is crucial for potential applications.

\subsection{Some explicit examples}
\label{sec:explicit_examples}
The complete  Hilbert space of our system of spins on $L_x \times L_y$ lattice has $4^{\cal N}$ dimensions, ${\cal N}=L_x L_y$. States are represented by  configurations 
\eq
\{i_1,i_2,\dots, i_{\cN} \}. \label{sts}
\eqx
of ${\cN}$ Dirac indices, $i_n=1,\dots,4$ with $n=1,\dots,\cN$ labelling sites of the lattice. All operators are constructed from tensor products 
of ${\cal N}$-fold four dimensional gamma matrices and the unity.\footnote{We use the specific representation of $\Gamma^k$ (cf. Table \ref{tab:gammas}), any other equivalent choice is possible.} In principle they require $\left(4^{\cal N}\right)^2$ units of computer storage, however in general they are sparse matrices and take only $O \left( 4^{\cal N} \right)$ memory size. Still, the memory requirement is the main limitation for such a direct approach and restricts available sizes to ca.   $ \cN  \sim 16 $.

 \begin{table}[h]
 \begin{center}
 \begin{tabular}{c c c c c}
 $\Gamma^1$      & $\Gamma^2$  & $\Gamma^3$ & $\Gamma^4$ & $\Gamma^5$ \\ 
 \hline\hline \\
       $ \left( \begin{array}{rrrr}
0 & 0 & -1 & 0\\
0 & 0 & 0 & 1\\
-1 & 0 & 0 & 0\\
0 & 1 & 0 & 0
\end{array}
\right)$ 
                 & $ \left( \begin{array}{rrrr}
0& 1 & 0 & 0\\
1& 0 & 0 & 0\\
0 & 0 & 0 & 1\\
0 & 0 & 1 & 0
\end{array}
\right) $    &  $ \left( \begin{array}{rrrr}
0& -i & 0 & 0\\
i& 0 & 0 & 0\\
0 & 0 & 0 & -i\\
0 & 0 & i & 0
\end{array}
\right) $    &   $ \left( \begin{array}{rrrr}
0 & 0 & -i & 0\\
0 & 0 & 0 & 1\\
i & 0 & 0 & 0\\
0 & -i & 0 & 0
\end{array}
\right) $             & $ \left( \begin{array}{rrrr}
1 & 0 & 0 & 0\\
0 & -1 & 0 & 0\\
0& 0 & -1 & 0\\
0 & 0 & 0 & 1
\end{array}
\right)  $\\
  \end{tabular}
  \caption {Explicit representation of  euclidean Dirac matrices used in this Section.} 
  \label{tab:gammas}
  \end{center}
  \end{table}

To reduce further the memory demand, we split the whole Hilbert space into $\cN+1$ sectors of the fixed fermion multiplicity (eigenvalue of $N$) $p=0,1,\dots,\cN$. In the fermionic representation the total number of fermions is obviously conserved. The same is true in our spin representation. Namely, the corresponding number operator
\eq
N=\sum_n \frac{1}{2}\left(1-\eta \Gamma^5(n)\right),
\eqx
commutes with the hamiltonian \eqref{hs}. Moreover, it also commutes with all plaquette and line operators. This allows to carry out the analysis of constraints in the sectors of fixed $p$ independently. Choosing the sector of fixed multiplicity amounts to restricting the full basis to states \eqref{sts} with $\cN - p$ indices $i$ in the "vacuum class", i.e.~$i=2$ or $3$; then remaining $p$ indices $i'$ are in the "excitation class", $i'=1$ or $4$. 

%With our choice of gamma matrices and $\eta=-1$, $a=(2,3)$ and $b=(1,4)$.
In practical terms, we will now be dealing with the $\cN+1$ fixed multiplicity sectors of the full Hilbert space separately, the size of each sector being 
\eq
2^{\cN} \left( \begin{array}{c}
                                                  \cN \\
                                                  p
                                       \end{array}
                  \right)  \longrightarrow \left( \begin{array}{c}
                                                  \cN \\
                                                  p
                                       \end{array}
                  \right), \label{red1}
\eqx
before and after imposing constraints in the spin representation.

Moreover, constraint operators commute not only with the number operator $N$ but also with each of the individual densities $N(n)$. This allows to further split the problem by performing the reduction of Hilbert space in each sub-sector of fixed $p$ {\em and} fixed  positions of $p$ spin excitations $r_1,r_2,\dots,r_p$ (or equivalently, fermionic coordinates), in the configuration space. Now the reduction of dimension takes the form
\eq
2^{\cN}   \longrightarrow  1.  \label{red2}
\eqx
Restriction to subspaces with fixed eignenvalues of $N(n)$ allows to save computer memory. Furthermore solutions of constraints obtained this way have clear physical interpretation, as they are parametrized by space coordinates of $p$ fermions. This  is valid for all lattice sizes.  It should be noted, however, that  reduction \eqref{red2} is possible only for the purpose of studying the constraints. The reduced spin hamiltonian has to be calculated in the bigger subspace of fixed $p$. The basis of this subspace, consisting of $ \left( \begin{array}{c}
                                                  \cN \\
                                                  p
                                       \end{array}
                  \right)$ vectors obtained by performing the reduction \eqref{red2} separately for each of $ \left( \begin{array}{c}
                                                  \cN \\
                                                  p
                                       \end{array}
                  \right)$ possible density configurations. This provides an appropriate basis of {\em constraint-satisfying} spin excitations in the larger sector of fixed fermionic multiplicity $p$.
 
To proceed further, we define the projection operators associated with all plaquettes and two Polyakov lines
 
 \eq
 \Sigma_{n}=\frac{1}{2}(1+P_{n})\;\;\;\;\;\;\;\Sigma_Z=\frac{1}{2}(1+{\cal L}_Z),\;\;Z=x,y,
 \eqx
 and calculate their matrix representations, at fixed total multiplicity $p$. For illustration we explicitly display below traces of successive products of all relevant projectors on $3\times3$ and $4\times4$ lattices. 

\begin{table}[h!]
\begin{center}
\begin{tabular}{|r|c|c|c|c|c|c|c|c|c|c|}
\hline
\textbf{p=}&\textbf{0}&\textbf{1}&\textbf{2}&\textbf{3}&\textbf{4}&\textbf{5}&\textbf{6}&\textbf{7}&\textbf{8}&\textbf{9}\\
\hline
$\text{Tr }\Sigma_{11}$&256&2304&9216&21504&32256&32256&21504&9216&2304&256\\
\hline
$\text{Tr }\Sigma_{11}\Sigma_{12}$&128&1152&4608&10752&16128&16128&10752&4608&1152&128\\
\hline
$\text{Tr }\Sigma_{11}\Sigma_{12}\Sigma_{13}$&64&576&2304&5376&8064&8064&5376&2304&576&64\\
\hline
$\text{Tr }\Sigma_{11}\Sigma_{12}...\Sigma_{21}$&32&288&1152&2688&4032&4032&2688&1152&288&32\\
\hline
$\text{Tr }\Sigma_{11}\Sigma_{12}...\Sigma_{22}$&16&144&576&1344&2016&2016&1344&576&144&16\\
\hline
$\text{Tr }\Sigma_{11}\Sigma_{12}...\Sigma_{23}$&8&72&288&672&1008&1008&672&288&72&8\\
\hline
$\text{Tr }\Sigma_{11}\Sigma_{12}...\Sigma_{31}$&4&36&144&336&504&504&336&144&36&4\\
\hline
$\text{Tr }\Sigma_{11}\Sigma_{12}...\Sigma_{32}$&2&18&72&168&252&252&168&72&18&2\\
\hline
$\text{Tr }\Sigma_{11}\Sigma_{12}...\Sigma_{33}$&2&18&72&168&252&252&168&72&18&2\\
\hline
$\text{Tr }\Sigma_{11}\Sigma_{12}...\Sigma_{x}$&1&9&36&84&126&126&84&36&9&1\\
\hline
$\text{Tr }\Sigma_{11}\Sigma_{12}...\Sigma_{y}$&0&9&0&84&0&126&0&36&0&1\\
\hline
\end{tabular}
\caption{Reduction of the spin Hilbert space for $3\times3$ lattice in $p$-particle sectors. Periodic boundary conditions are assumed.}
\label{tab:3times3}
\end{center}
\end{table}

For $3\times3$ lattice (Table \ref{tab:3times3}) the reduction was performed in sectors of fixed fermion multiplicity $p$ and proceeds according to the scheme \eqref{red1}. Indeed, including successive projectors reduces dimensions by half, as expected. The last (here $\Sigma_{33}$) plaquette projector does not further reduce the dimension, in agreement with the earlier discussion. Moreover, the final result is nonzero only for multiplicities which satisfy \eqref{cond}. Finally, the second Polyakov line is dependent on other projectors, as is $\Sigma_{33}$, for allowed multiplicities, while it is incompatible with the rest for forbidden values of $p$. The final dimensionalities of the fully reduced spin Hilbert spaces agree with the sizes of the corresponding sectors with $p$ indistinguishable fermions \eqref{red1}, as it should be.

\begin{table}[h!]
\begin{center}
\begin{tabular}{|c|c|c|c|}
\hline
\multicolumn{2}{|c|}{Sector ($p$)}&even, $0\leq p\leq 16$&odd, $0<p<16$\\
\hline
\multicolumn{2}{|c|}{Occupied sites}&\multicolumn{2}{|c|}{from \# 1 to \# p}\\
\hline
\multicolumn{1}{|c|}{\multirow{18}{*}{
\parbox[t]{2mm}{\multirow{3}{*}{\rotatebox[origin=c]{90}{Hilbert space reduction}}}
}}&\multicolumn{1}{|c|}{Tr $\Sigma_{11}$}&\multicolumn{2}{|c|}{32768}\\ \cline{2-4}
\multicolumn{1}{|c|}{}&\multicolumn{1}{|c|}{Tr $\Sigma_{11}\Sigma_{21}$}&\multicolumn{2}{|c|}{16384}\\ \cline{2-4}
\multicolumn{1}{|c|}{}&\multicolumn{1}{|c|}{Tr $\Sigma_{11}...\Sigma_{31}$}&\multicolumn{2}{|c|}{8192}\\ \cline{2-4}
\multicolumn{1}{|c|}{}&\multicolumn{1}{|c|}{Tr $\Sigma_{11}...\Sigma_{41}$}&\multicolumn{2}{|c|}{4096}\\ \cline{2-4}
\multicolumn{1}{|c|}{}&\multicolumn{1}{|c|}{Tr $\Sigma_{11}...\Sigma_{12}$}&\multicolumn{2}{|c|}{2048}\\ \cline{2-4}
\multicolumn{1}{|c|}{}&\multicolumn{1}{|c|}{Tr $\Sigma_{11}...\Sigma_{22}$}&\multicolumn{2}{|c|}{1024}\\ \cline{2-4}
\multicolumn{1}{|c|}{}&\multicolumn{1}{|c|}{Tr $\Sigma_{11}...\Sigma_{32}$}&\multicolumn{2}{|c|}{512}\\ \cline{2-4}
\multicolumn{1}{|c|}{}&\multicolumn{1}{|c|}{Tr $\Sigma_{11}...\Sigma_{42}$}&\multicolumn{2}{|c|}{256}\\ \cline{2-4}
\multicolumn{1}{|c|}{}&\multicolumn{1}{|c|}{Tr $\Sigma_{11}...\Sigma_{13}$}&\multicolumn{2}{|c|}{128}\\ \cline{2-4}
\multicolumn{1}{|c|}{}&\multicolumn{1}{|c|}{Tr $\Sigma_{11}...\Sigma_{23}$}&\multicolumn{2}{|c|}{64}\\ \cline{2-4}
\multicolumn{1}{|c|}{}&\multicolumn{1}{|c|}{Tr $\Sigma_{11}...\Sigma_{33}$}&\multicolumn{2}{|c|}{32}\\ \cline{2-4}
\multicolumn{1}{|c|}{}&\multicolumn{1}{|c|}{Tr $\Sigma_{11}...\Sigma_{43}$}&\multicolumn{2}{|c|}{16}\\ \cline{2-4}
\multicolumn{1}{|c|}{}&\multicolumn{1}{|c|}{Tr $\Sigma_{11}...\Sigma_{14}$}&\multicolumn{2}{|c|}{8}\\ \cline{2-4}
\multicolumn{1}{|c|}{}&\multicolumn{1}{|c|}{Tr $\Sigma_{11}...\Sigma_{24}$}&\multicolumn{2}{|c|}{4}\\ \cline{2-4}
\multicolumn{1}{|c|}{}&\multicolumn{1}{|c|}{Tr $\Sigma_{11}...\Sigma_{x}$}&\multicolumn{2}{|c|}{2}\\ \cline{2-4}
\multicolumn{1}{|c|}{}&\multicolumn{1}{|c|}{Tr $\Sigma_{11}...\Sigma_{y}$}&\multicolumn{2}{|c|}{1}\\ \cline{2-4}
\multicolumn{1}{|c|}{}&\multicolumn{1}{|c|}{Tr $\Sigma_{11}...\Sigma_{34}$}&1&0\\ \cline{2-4}
\multicolumn{1}{|c|}{}&\multicolumn{1}{|c|}{Tr $\Sigma_{11}...\Sigma_{44}$}&1&0\\
 \hline
\end{tabular}
\caption{Reduction of the spin Hilbert space for subsectors $0\leq p\leq 16$, and fixed coordinates, on a $4\times 4$ lattice. Sites of the lattice are ordered lexicographically, thus e.g. sites from \#1 to \#5 means sites $(1,1)$, $(2,1)$, $(3,1)$, $(4,1)$ and $(1,2)$.}
\label{tab:4times4}
\end{center}
\end{table}

In the $4\times4$ case the reduction was done in subsectors of fixed $p$ fermionic coordinates (scheme \eqref{red2}). Each of these has the same dimension $2^{\mathcal N}$, independently of $p$.  As in the previous case, adding subsequent plaquette projectors reduces the size by half until one reaches the last two plaquettes. Interestingly, neither of these further reduces the remaining Hilbert space. This means that for $4 \times 4$ lattice (and more generally for (even)$\times$(even) ones) two plaquettes are dependent. This is easy to explain: for even-by-even lattices one can split all plaquettes into two classes, according to the value of $(-1)^{n_x + n_y}$, where $n = (n_x, n_y)$ is the coordinate of the lower-left corner of the plaquette. Then for each of the two groups independently one has the relation 
\eq 
\prod\limits_{n} P_{n}=(-1)^N, \qquad n_x+n_y \;\;even  \text{ or } odd.
\eqx 
 Consequently, on (even)$\times$(even) lattices two plaquette projectors can be expressed in terms of the other. This explains the content of the Table \ref{tab:4times4}.

On the other hand both Polyakov line projectors are now independent. This has been explained in the discussion below equation \eqref{prod}. Regardless of parities of $L_x$ and $L_y$ the number of independent projectors is $\mathcal N$, although they are distributed in a different way between plaquette and line operators.

The whole discussion can be repeated for other situations as well. The results are summarized in Table \ref{tab:summary} for all four cases.

  \begin{table}[h]
 \begin{center}
 \begin{tabular}{c c c c c} \hline\hline
 $L_x$      & $L_y$  & plaquettes & lines & multiplicity \\ \hline
 odd          & odd       &  $\cN -1$    &   ${\cal L}_x$  or ${\cal L}_y$           & odd \\
 odd          & even     &  $\cN -1$     & ${\cal L}_x$ & odd \\
 even        & odd       &  $\cN -1$     & ${\cal L}_y$ & odd \\
 even        & even     &  $\cN -2$   &      ${\cal L}_x$  and ${\cal L}_y$            & even \\ \hline
  \end{tabular}
  \caption {Number of independent projectors and consistent multiplicities for periodic boundary conditions in both representations, $\epsilon=\epsilon'=1$.} 
  \label{tab:summary}
  \end{center}
  \end{table}

The final consistency check is to calculate the spectrum of the spin hamiltonian in the subspace defined by the constraints. Using methods outlined above we construct for each $p$ a basis of states satisfying all constraints. For small lattices considered in this example (see also the next Section), all eigenvectors of combined projectors are analytically generated by Mathematica \cite{MATH}. Having done that, matrix elements of the reduced spin hamiltonian in the relevant subspace can be calculated. This exercise has been repeated for several multiplicity sectors on above lattices. In each of the considered cases, the complete spectrum of known eigenenergies of $p$ free fermions was analytically reproduced.

\section{Modified constraints and background fields}
\label{sec:generalization}
Above discussion addressed solely the case where all plaquette operators were constrained to unity.
In principle, however, one could consider the whole family of $2^{\cN}$ modified constraints
\eq
P_{n}=\pm 1, \;\;\;\;\;\;\;\;\;  1 \leqslant n \leqslant  \cN.
\eqx
Such sectors obviously exist in the unconstrained spin system, which raises the question of their interpretation. The answer is simple and instructive, as will be discussed now. 

Consider the following modification of the original fermionic hamiltonian \eqref{hf}
\eqn
H_f&=& i \sum_{\vn,\ve}\left(U(\vn, \vn+\ve)\phi(\vn)^{\dagger}\phi(\vn+\ve) - U(\vn, \vn+\ve)\phi(\vn+\ve)^{\dagger}\phi(\vn)\right)\\ 
&=&\frac{1}{2} \sum_l  \left(U(l) S(l) + U(l) \widetilde{S}(l)\right),  \label{hfU} 
\eqnx
where $U(l)$ is an additional $\mathbb{Z}_2$ field assigned to links $l$. Then in the spin representation
\eq
H_s=\frac{1}{2} \sum_l  \left(U(l) S(l) + U(l) \widetilde{S}(l)\right).   \label{hsU}
\eqx
with the same variables $U(l)$, and $S(l)$ given by \eqref{lgam}. Clearly these hamiltonians describe fermions and/or corresponding spins in a background $\mathbb{Z}_2$ gauge field. As for free hamiltonian (and~more generally any hamiltonian), systems described by $H_f$ and $H_s$ are equivalent, as~long as we restrict the spin Hilbert space in a way discussed in the previous Section. We note in passing that this provides an extension of the fermion-spin equivalence for the case of external fields as well.

Interestingly, it is also possible to introduce the background gauge field in such a way that it is not explicitly visible in the spin hamiltonian.\footnote{An early version of this observation was made already in \cite{SZ}.} Indeed, one can absorb the $U(l)$ factors into new hopping operators\footnote{From the gauge theory perspective these are the gauge covariant hopping operators.} and define 
\eq
S'(l)=U(l) S(l),\qquad\widetilde{S}'(l)=U(l) \widetilde{S}(l).
\eqx
This does not change the commutation rules obeyed by these operators. Now the spin hamiltonian does not explicitly depend on the external field 
\eq
H_s'=\frac{1}{2} \sum_l   \left(S'(l) +  \widetilde{S}'(l)\right),   \label{hsp}
\eqx
but the constraints on the new spin variables do. They readily follow from \eqref{cons}
\eq
 P'_n=\prod_{l  \in C_n }U(l). \label{consX}
 \eqx
That is, the system of new spins is not free, but remembers the interactions via constraints \eqref{consX} only. In other words, there are two ways of introducing minimal interaction with the external field: 
\begin{enumerate}
    \item by introducing link variables explicitly into the hamiltonian and imposing the "free" form of the constraint \eqref{con},
    \item by using the free spin hamiltonian \eqref{hs} with "interacting" constraint \eqref{consX}.
\end{enumerate}
We emphasize that the first method is viable for any interactions, because the equivalence between fermions and spins is valid for any hamiltonian. The second method is possible due to the specific structure of the minimal coupling, which amounts to introducing parallel transports in any term in the hamiltonian which involves products of on distinct lattice sites charged under the gauged symmetry. It provides an interesting interpretation of the whole spin Hilbert space. 

On the fermionic side, the hamiltonian \eqref{hfU} is that of two dimensional fermions in the fixed, external gauge field of the Wegner type \cite{WG}. The gauge field is not dynamical. On~the other hand, our spin system is also coupled to the same gauge field: various boundary conditions are probing different gauge invariant classes of the $\mathbb{Z}_2$ variables \cite{FS}.

The phenomenon discussed above will be illustrated by working out a simple example in Subsection \ref{sec:soluble_example}.

One particularly interesting feature of the presented construction is that the allowed value of $(-1)^N$ becomes dependent on the background field. More precisely, let $(-1)^{N_0}$ be the right hand side of \eqref{cond}. In presence of the field $U(l)$, relation \eqref{cond} is modified to
\begin{equation}
(-1)^N = (-1)^{N_0} \cdot \prod_{l} U(l),
\end{equation}
where the product is taken over all links of the lattice. Derivation of this formula is analogous to the case of vanishing background field. An interesting gauge-theoretic interpretation of this relation has been proposed in \cite{BBR} and is briefly reviewed in Section \ref{sec:summary}.
 
\subsection{A soluble example}
\label{sec:soluble_example}
Consider the configuration of Wegner variables given by
\eq
U_x(x,y)=(-1)^y,\;\;\;\; U_y(x,y)=1,   \label{MU}
\eqx
where we assume that $L_y$ is even. In this case the fermionic hamiltonian \eqref{hfU} can be diagonalized analytically.\footnote{One employs discrete Fourier transformation and a Bogoliubov transformation.} The one-particle spectrum reads
\eq
E^{(1)}_{\mathrm{mag}}(k_x,k_y)=\pm 2  \sqrt{\sin^2{\left(\frac{2\pi k_x}{L_x}\right)} +  \sin^2{\left(\frac{2\pi k_y}{L_y}\right)}},\;\;\;\; 
1 \leqslant k_x \leqslant L_x,\;\;\;1 \leqslant  k_y \leqslant  \frac{L_y}{2},
\eqx
while in the free case one has
\eq
E^{(1)}_{\mathrm{free}}(k_x,k_y)=2  \sin{\left(\frac{2\pi k_x}{L_x}\right)} + 2 \sin{\left(\frac{2\pi k_y}{L_y}\right)},\;\;\;\; 1 \leqslant k_z \leqslant L_z, \;\;\;z=x,y.
\eqx
Configuration \eqref{MU} results in all plaquettes being equal
\eq
P_{n}=-1,\;\;\;\; 1 \leqslant n \leqslant \cN,  \label{PM}
\eqx
so it is a Wegner's version of a constant magnetic field. 

We have repeated the procedure outlined in Section \ref{sec:explicit_examples} for the $3\times4$ lattice in order to reproduce this result. Table \ref{tab:3times4} shows, familiar by now, pattern of reduction of Hilbert spaces. All proceeds as before, the new element being the distinguished role of the line projector associated with ${\cal L}_x$, as presented in Table \ref{tab:summary}. 

 \begin{table}[]
\begin{center}
\begin{tabular}{|c|c|c|c|c|c|}
\hline
\multicolumn{2}{|c|}{\textbf{p}}&\multicolumn{4}{c|}{\bf{1}}\\
\hline
\multicolumn{2}{|c|}{$\text{Tr 1}$} & \multicolumn{4}{c|}{49152}     \\
\hline
\multicolumn{2}{|c|}{$\text{Tr }\Sigma_{11}$}  & \multicolumn{4}{c|}{24576}     \\
\hline
\multicolumn{2}{|c|}{$\text{Tr }\Sigma_{11}\Sigma_{21}$}  & \multicolumn{4}{c|}{12288}     \\
\hline
\multicolumn{2}{|c|}{$\text{Tr }\dots\Sigma_{31}$}  & \multicolumn{4}{c|}{6144}     \\
\hline
\multicolumn{2}{|c|}{$\text{Tr }\dots\Sigma_{12}$}  & \multicolumn{4}{c|}{3072}     \\
\hline
\multicolumn{2}{|c|}{$\text{Tr }\dots\Sigma_{22}$}  & \multicolumn{4}{c|}{1536}     \\
\hline
\multicolumn{2}{|c|}{$\text{Tr }\dots\Sigma_{32}$}  & \multicolumn{4}{c|}{768}     \\
\hline
\multicolumn{2}{|c|}{$\text{Tr }\dots\Sigma_{13}$}  & \multicolumn{4}{c|}{348}     \\
\hline
\multicolumn{2}{|c|}{$\text{Tr }\dots\Sigma_{23}$}  & \multicolumn{4}{c|}{192}     \\
\hline
\multicolumn{2}{|c|}{$\text{Tr }\dots\Sigma_{33}$}  & \multicolumn{4}{c|}{96}     \\
\hline
$\text{Tr }\dots\Sigma_{14}  $ & 48  & $\text{Tr }\dots\Sigma_{14}  $ & 48 &   $\text{Tr }\dots\Sigma_{x}  $  &  48       \\
\hline
$\text{Tr }\dots\Sigma_{24}  $ & 24  & $\text{Tr }\dots\Sigma_{24}  $ & 24 &  $\text{Tr }\dots\Sigma_{y}  $   &   24      \\
\hline
$\text{Tr }\dots\Sigma_{34}  $ & 24  & $\text{Tr }\dots\Sigma_{34}  $ & 24 &  $\text{Tr }\dots\Sigma_{14}  $   &   12     \\
\hline
$\text{Tr }\dots\Sigma_x  $ & 12  & $\text{Tr }\dots\Sigma_y  $ & 24 &  $\text{Tr }\dots\Sigma_{24} $   &      12  \\
\hline
$\text{Tr }\dots\Sigma_y  $ & 12  & $\text{Tr }\dots\Sigma_x  $ & 12 &   $\text{Tr }\dots\Sigma_{34}  $   &     12  \\
\hline
\multicolumn{2}{|c|}{\bf A}  & \multicolumn{2}{|c|}{\bf B} & \multicolumn{2}{|c|}{\bf C}        \\
\hline
\end{tabular}
\caption{Reduction of the spin Hilbert space for $3\times4$ lattice in the one excitation  sector, and with different ordering ($A,B,C$) of projectors. Periodic boundary conditions are used.}
\label{tab:3times4}
\end{center}
\end{table}

Table \ref{tab:3times4} displays results for three different orderings ($A, B, C$) of applying projectors. Although the final effect is the same\footnote{And again consistent with the condition \eqref{cond}.}, the results in the intermediate stages are different, as will be explained now. Orderings $A$ and $B$ differ only by the order of the two line projectors which are added at the end of the process. Before that, we employ all $\cN=12$ plaquette projectors. As discussed before, the last one is dependent on the rest. Then, among the two line projectors, $\Sigma_y$ is ineffective, i.e. dependent on other projectors, while $\Sigma_x$ is independent and reduces the remaining space, regardless of the ordering $A$ or $B$ of imposing the constraints. 

The situation is different in the scheme $C$, in which line projectors are imposed before the last three plaquettes. In this case $\Sigma_y$ acts as an independent projector. This does not contradict the discussion below equation \eqref{prod}, because operators $\mathcal L_y(n_x)$ are independent for different $n_x$ if not all plaquette constraints are imposed (indeed, their ratio is precisely the product of some number of plaquette operators). The total number of independent constraints is equal to $\mathcal N$, so two among the last three plaquette constraints in the ordering $C$ have to be ineffective. This is indeed seen in the Table \ref{tab:3times4}. Furthermore the final size of the one particle sector is the correct one.  

Matrix elements of the spin hamiltonian in the constrained one particle sector was calculated for two choices of boundary conditions:
\begin{enumerate}
    \item free \eqref{con} together with ${\cal L}_x(1)=1$, ${\cal L}_y(1)=1$,
    \item magnetic \eqref{PM} and ${\cal L}_x(1)=-1$, ${\cal L}_y(1)=1$.
\end{enumerate}

In both cases correct fermionic spectrum was reproduced from the reduced spin hamiltonian. 

\section{Summary and outline}
\label{sec:summary}

An old proposal for local bosonization of fermionic degrees of freedom in general dimensions was revisited. Resulting spin systems are indeed local. They are subject to additional constraints which, even though local themselves, introduce effectively long range interactions. In particular they are sensitive to the lattice geometry and fermionic multiplicities.

In this paper we have studied and classified this dependence in detail. The necessary reduction of spin Hilbert space was demonstrated analytically for several small lattices. A~number of regularities has been found. We have provided explanations which are valid for larger systems as well. Most importantly, for a given lattice size and boundary conditions, the fermion-spin equivalence holds only in the subspace defined by one of the two possible values of the fermionic parity. In this sector imposing all constraints resulted in reduction of the spin Hilbert space to dimension appropriate for fermions.

%In these sectors complete reduction to the correct fermionic Hilbert space could be achieved. The general analytic  conditions when this occurs were derived.

For the above small lattices all relevant constraints were solved with the aid of Mathematica. Consequently, complete eigenbases of spin states fulfilling the constraints are known analytically. Their structure is tantalizingly simple. Explicit generalization to arbitrary lattice sizes still remains a challenge.

The second step was to calculate the spectra of proposed spin hamiltonians, reduced to the subspace defined by constraints. In all considered cases the well known fermionic eigenenergies have been readily reproduced.

Afterwards, the equivalence was generalized to fermions coupled minimally to a background $\mathbb{Z}_2$ gauge field. Apart from being interesting by itself, this provided a simple and intuitive interpretation of the constraints: changing the value of constraint operators is equivalent to coupling fermions to the background field. This can be achieved without introducing the background field explicitly in the spin hamiltonian. All constraints, conceivable for this system, split into gauge invariant classes which, are in one to one correspondence with all possible gauge orbits of the external $\mathbb Z_2$ field. A simple proof of this fact was given. In addition, the consistency of the whole scheme was directly checked for a particular configuration of $\mathbb Z_2$ variables -  the Wegner's analog of a~constant magnetic field. Indeed, the analytically obtained spectrum of the spin hamiltonian, reduced to the constraint-fulfilling sector, reproduced the fermionic eigenenergies in this field.

Summarizing, the exact equivalence between lattice fermions and constrained Ising-like spins was checked for a range of small lattices in ($2+1$) dimensions. The interplay between the  constraints, lattice geometry and boundary conditions is now fully understood and classified for all fermion multiplicities and all lattice volumes. Moreover, for above small systems the constraints were explicitly solved leading to the direct construction of the reduced spin Hilbert spaces.  From a~practitioner's standpoint this provides convincing evidence for the validity of the fermion-spin equivalence by itself, since one would generally not expect an \textit{exact} duality to hold by accident and only for small lattice sizes \footnote{In other words: nothing qualitatively dramatic happens with increasing lattice size. Even signatures of such subtle phenomena as phase transitions build up gradually with increasing the volume. }. Proofs of validity for quite general lattice geometries and arbitrary volumes are now available in the literature, but until now almost no practical implementations have been presented. This gap is now filled. 

%It is a safe assumption, that increasing lattice size does not change qualitatively this result. It is well known that no dramatic effects occur as lattice sizes are progressively increased. Even such subtle phenomena as phase transitions set up gradually with increasing the volume of the system. The best known example is provided by the specific heat of the Ising model $c_L(T)$. Already for lattices as small as $3 \times 3$ $c_L(T)$ develops a broad maximum in $T$ which systematically shrinks, shifts and eventually turns into a singularity at infinite volume. In our case, once the major small volume effects, like correlations between the fermion number and lattice geometry, together with bounds between various constraints, have been understood, nothing dramatic happens on the road to the infinite volume. Consequently we claim that the fermi-spin equivalence proposed in this paper, is very plausible at all volumes, even if not rigorously proven. 

For simplicity, most of the discussion and our calculations concentrated on the two dimensional case. Nevertheless, extension to higher space dimensions does not present any conceptual difficulties, and in fact does not bring any qualitatively new theoretical features.

Numerous dualities between various ($2+1$) dimensional theories have been recently discovered (for reviews and references see e.g. \cite{KT,SSWX}). Building on the seminal papers of Peskin, Polyakov and others \cite{MP,AMP,DH}, there was a steady growth of understanding of various phenomena \cite{FW,AMP2,JK,KLNRR}. This culminated in a dramatic increase of interest in the subject in the last few years \cite{SSWW,KT2,DTS,CSWR,OA}. Many new structures have been found even behind the simplest and classic by now, Kramers-Wannier duality in ($1+1$) dimensions \cite{SSWX,KT3}. 
To~our knowledge, however, none of the available up to date dualities accounts exactly for the bosonization studied in this paper. On the other hand there are several structural similarities, which we point out below.

Since gamma matrices employed here can be represented as tensor products of two Pauli matrices, our bosonization connects free fermions to a system roughly viewed as {\em pairs} of Ising spins living at lattice sites. Upon imposing constraints such a model becomes exactly equivalent to above fermions \eqref{hf2}. A nontrivial relation emerges between the value of conserved
$\mathbb Z_2$ charge $(-1)^N$ on one side of the duality and boundary conditions on the other. Such phenomena occur already for dualities as those of Jordan and Wigner or Kramers and Wannier, as can be seen upon carefully keeping track of various signs and global constraints, see \cite{R} for a~detailed review.

Alternatively, the unconstrained pairs of spins with local Ising-like interactions should describe fermions interacting with a dynamical $\mathbb{Z}_2$ field. An attempt to construct such a theory was recently reported in \cite{BBR}.

Most of dualities mentioned above involve some dynamical gauge field $A$. It is often the case that this gauge field obeys a modified form of the Gauss' law\footnote{Here we have in mind the hamiltonian formalism with temporal gauge.}, which involves field-dependent phase factors. This is related to the fact that the gauge field action is not exactly gauge invariant, but its gauge variation depends only on its value on the spacetime boundary, and hence can be absorbed into a redefinition of the initial and final state wave functions. Such mechanism is at work in particular in Chern-Simons theories and their version suitable for finite groups, introduced by Dijkgraaf and Witten \cite{DW}. Modification of the Gauss' law has the consequence that magnetic flux excitations become paired with electric charges. This mechanism, known as flux attachment, may lead to a transmutation of statistics, due to presence of Aharonov-Bohm phases \cite{FW, SSWW}. Interestingly, the $\mathbb{Z}_2$ gauge field introduced in Section \ref{sec:generalization} does also have these properties \cite{BBR,BR}.

Mapping presented here is an exact relation between microscopic degrees of freedom for fermions and spins, as in the Jordan-Wigner duality \cite{SSWX}. This is different than some of the recently proposed dualities, which connect effective theories in vicinities of RG fixed points. These are typically very difficult to establish rigorously. However, one can still make arguments based on universality, matching of symmetries and anomalies, etc. 

It is an attractive possibility that results established in this paper provide a microscopic realization of one of the "web of dualities" discussed e.g. in \cite{KT,SSWW}. One possible candidate would be the duality between a scalar field and a fermi-gauge system described in \cite{KT}. We~are looking forward to study some of these questions in detail.

Finally, we remark that bosonization discussed in this work can be extended to higher dimensions simply by using higher dimensional Clifford algebras. In $d$ space dimensions this would lead to a $d$-plet of Ising spins living at each lattice site and interacting with nearest-neighbour couplings. It would then be interesting to see if such a mapping has its counterpart among the recently proposed webs of dualities.

\acknowledgments

We thank the unknown referee for drawing our attention to the recent developments in ($2+1$) dimensional dualities.
This work is supported in part by the NCN grant: UMO-2016/21/B/ST2/01492. BR was also supported by the MNS donation for PhD students and young scientists N17/MNS/000040.

\end{document}